\documentclass[preprintnumbers,nofootinbib,superscriptaddress,10pt]{revtex4}
\usepackage[dvipdfmx]{graphicx} 
\usepackage{amsmath,amssymb,amsfonts, bm} 

\def\a{\alpha}    \def\d{\delta} \def\D{\Delta} \def\e{\epsilon}   \def\th{\theta}     \def\m{\mu} \def\n{\nu}      \def\s{\sigma}  \def\t{\tau}       

\def\dg{\dagger}  \def\nn{\nonumber}

\newcommand{\meV}{ {\rm meV} }

    \newcommand{\To}{\Rightarrow}

\def\abs#1{\left| #1\right|}

\newcommand{\Diag}[3]{ \begin{pmatrix} #1 & 0 & 0 \\ 0 & #2 & 0 \\ 0 & 0 & #3 \\\end{pmatrix}}

\usepackage{color}

\begin{document}

\title{\large Analytic structure of the MNS matrix with diagonal reflection symmetries and its physical consequences}

\preprint{STUPP-22-256}
\author{Masaki J. S. Yang}
\email{mjsyang@mail.saitama-u.ac.jp}
\affiliation{Department of Physics, Saitama University, 
Shimo-okubo, Sakura-ku, Saitama, 338-8570, Japan}



\begin{abstract} 

In this paper, we survey the analytic structure of the MNS matrix with diagonal reflection symmetries. 
If the mass matrix of charged leptons $m_{e}$ is hierarchical (i.e., $|(m_{e})_{33}| \simeq m_{\t} \gg |(m_{e})_{1i , j1}|$), by neglecting the 1-3 mixing of $m_{e}$, the MNS matrix is represented by four parameters and several sign degrees of freedom.
By substituting the three observed mixing angles $\theta_{ij}$ as input parameters, 
the Dirac phase $\delta$  and the Majorana phases $\alpha_{2,3}$ are represented by functions of the 1-2 mixing of charged leptons $s_{e}$. 
The effective mass of double beta decay $m_{ee}$ is also displayed as a function of $s_{e}$ and 
the lightest neutrino mass $m_{1 \, \rm or \, 3}$.
Because the generalized CP symmetries restrict the effective mass to near the maximum or minimum value in the whole parameter region, several scenarios are suggested to be excluded by the latest limit of the KAMLAND-Zen collaboration. 

\end{abstract} 

\maketitle

\section{Introduction}

The Dirac phase $\d$, which represents CP violation (CPV) in the lepton mixing matrix, has been measured recently \cite{T2K:2021xwb, NOvA:2021nfi}. 
Since the leptonic CPV will be determined more precisely by next-generation experiments \cite{Hyper-KamiokandeProto-:2015xww, DUNE:2020jqi}, the behavior of the phase is an important subject to be considered in this era. 
While the Dirac phase has been studied generally \cite{Ge:2011qn, Ge:2011ih, Petcov:2014laa, Girardi:2014faa, Girardi:2015vha, Delgadillo:2018tza},   
a generalized CP symmetry (GCP) \cite{Ecker:1981wv, Ecker:1983hz, Gronau:1985sp, Ecker:1987qp,Neufeld:1987wa,Ferreira:2009wh,Feruglio:2012cw,Holthausen:2012dk,Ding:2013bpa,Girardi:2013sza,Nishi:2013jqa,Ding:2013hpa,Feruglio:2013hia,Chen:2014wxa,Ding:2014ora,Ding:2014hva,Chen:2014tpa,Li:2015jxa,Turner:2015uta, Rodejohann:2017lre, Penedo:2017vtf, Nath:2018fvw} can fix the phase. 
One notable example is the $\m-\t$ reflection symmetry 
\cite{Harrison:2002et,Grimus:2003yn, Grimus:2005jk, Farzan:2006vj, Joshipura:2007sf, Adhikary:2009kz, Joshipura:2009tg, Xing:2010ez, Ge:2010js, He:2011kn, Gupta:2011ct, Grimus:2012hu, He:2012yt, Joshipura:2015dsa, Xing:2015fdg, He:2015afa, Chen:2015siy, He:2015xha, Samanta:2017kce, Xing:2017cwb, Nishi:2018vlz, Nath:2018hjx, Sinha:2018xof, Huang:2018fog, Xing:2019edp, Pan:2019qcc, Chakraborty:2019rjc, Liao:2019qbb, Yang:2020qsa, Duarah:2020zjo, Zhao:2021dwc, Bao:2022kon}  
that predicts the maximal Dirac phase $\d = \pm \pi/2$. 

On the other hand, diagonal reflection symmetries (DRS) \cite{ Yang:2020goc, Yang:2021smh, Yang:2021xob, Yang:2022gcs} are GCPs that can predict relatively small $\d$ in a way that can unify quarks and leptons.
Up to the present, 
various models that satisfy (approximately) these symmetries have been discussed in the context of zero-textures \cite{Shin:1985cg, Shin:1985vi, Lehmann:1995br, Kang:1997uv, Mondragon:1998gy, Fritzsch:1999im, Fritzsch:2002ga, Xing:2003yj, Xing:2009eg, Barranco:2010we, Xing:2015sva, Fritzsch:2021ipb, Awasthi:2022nbi}. 
However, the behavior of the symmetries is not yet well understood. 
In this paper, to explore general properties, 
we survey the analytic structure and physical observables 
which results from the MNS matrix with DRS. 

This paper is organized as follows. 
The next section gives a review of DRS in the quark sector. 
In Sec.~3, we discuss a field theoretical realization of DRS by the type-I seesaw mechanism. 
In Sec.~4, a systematic analysis of the MNS matrix with DRS is performed. 
The final section is devoted to a summary. 

\section{Diagonal reflection symmetries for quarks}

In this section, we define the diagonal reflection symmetries (DRS), 
generalized CP symmetries associated with chiral symmetries of the first generation. 
First, a representation of the CKM matrix proposed by Fritzsch and Xing is \cite{Fritzsch:1997fw}, 
\begin{align}
 V_{\rm CKM} =  U_{u}^{\dg} U_{d} &= 
\begin{pmatrix}
c_{u} & s_{u} & 0 \\
 - s_{u} & c_{u} & 0 \\
 0 & 0 & 1 \\
\end{pmatrix} 
\begin{pmatrix}
e^{- i \phi}  & 0 & 0 \\
 0 & c_{q} & s_{q} \\
 0 & - s_{q} & c_{q} \\
\end{pmatrix}
\begin{pmatrix}
c_{d} & - s_{d} & 0 \\
 s_{d} & c_{d} & 0 \\
 0 & 0 & 1 \\
\end{pmatrix} \\ 
& = 
\begin{pmatrix}
s_{u} s_{d} c_{q} + c_{u} c_{d} e^{ - i \phi} & s_{u} c_{d} c_{q} - c_{u} s_{d} e^{- i \phi} & s_{u} s_{q} \\
c_{u} s_{d} c_{q} - s_{u} c_{d}  e^{ - i \phi} & c_{u} c_{d} c_{q} + s_{u} s_{d} e^{ - i \phi}  & c_{u} s_{q} \\
- s_{d} s_{q} & - c_{d} s_{q} & c_{q} 
\end{pmatrix} ,
\label{VCKM1} 
\end{align}
where $s_{\th} \equiv \sin \th , c_{\th} \equiv \cos \th, \phi$ is a CP-violating phase.  
The CKM matrix is represented by four parameters, 
and the following relationships exist between the three mixing angles $s_{f}$ and the PDG parameterization of $s_{ij}^{\rm CKM}$;
\begin{align}
t_{u} = \abs{V_{ub} \over V_{cb}} = {s_{13}^{\rm CKM} \over s_{23}^{\rm CKM} c_{13}^{\rm CKM}}\, , ~~ 
t_{d} = \abs{ V_{td} \over  V_{ts}} \, , ~~ 
c_{q} = c_{23}^{\rm CKM} c_{13}^{\rm CKM} \, ,
\label{tf}
\end{align}
where $t_{\th} \equiv \tan \th$. 
From the absolute value of the element $|V_{us}| = |s_{u} c_{d} c_{q} - c_{u} s_{d} e^{- i \phi}|$, the phase $\phi$ is found to be
\begin{align}
\cos \phi = {(s_{u} c_{d} c_{q})^{2} +  (c_{u} s_{d})^{2} - |V_{us}|^{2}  \over 2 s_{u} c_{d} c_{q} c_{u} s_{d}} \, . 
\end{align}
The recent best fit of observables \cite{Zyla:2020zbs} 
\begin{align}
s_{12}^{\rm CKM} &= 0.22650, ~~ 
s_{23}^{\rm CKM} = 0.04053, ~~ 
s_{13}^{\rm CKM} = 0.00361, ~~
\d^{\rm CKM} = 1.196, 
\end{align}
determines the four parameters $s_{u, q,d}$ and $\phi$ as
\begin{align}
s_{u} & =  0.0887, ~~ s_{d} =  0.2100, ~~ 
 s_{q} = 0.04069, ~~ \phi = 88.66^{\circ} . 
\label{FZparameters}
\end{align}
Fixing this phase to $\pi/2$ will only produce an error of about $10^{-5}$ in the CKM matrix elements \cite{Yang:2021smh}. 

Thus, we can interpret unitary matrices $U_{u,d}$ diagonalizing the mass matrix of quarks $m_{u,d}$ as
\begin{align}
U_{u} & =  
\begin{pmatrix}
+i  & 0 & 0 \\
 0 & c_{t} & s_{t} \\
 0 & - s_{t} & c_{t} \\
\end{pmatrix}
\begin{pmatrix}
c_{u} & -s_{u} & 0 \\
 s_{u} & c_{u} & 0 \\
 0 & 0 & 1 \\
\end{pmatrix} 
, ~~~
U_{d} = 
\begin{pmatrix}
1 & 0 & 0 \\
 0 & c_{b} & s_{b} \\
 0 & - s_{b} & c_{b} \\
\end{pmatrix}
\begin{pmatrix}
c_{d} & - s_{d} & 0 \\
 s_{d} & c_{d} & 0 \\
 0 & 0 & 1 \\
\end{pmatrix} , \label{Ud}
\end{align}
where $s_{q} = s_{b} c_{t} - c_{b} s_{t}$. 
One prediction emerges due to the assumption of the phase $\phi = \pi /2$; 
\begin{align}
J & = s_{u} s_{d} c_{u} c_{d} c_{q} s_{q}^{2} 
 = (c_{12} s_{12} s_{23} c_{23} c_{13}^{2} s_{13})^{\rm CKM} \sin \d^{\rm CKM} 
\, .
\end{align}
Comparing the third column of the CKM matrix we obtain  $s_{u} s_{q} = s_{13}^{\rm CKM}$ and $c_{u} s_{q} = s_{23}^{\rm CKM } c_{13} ^{\rm CKM} $.  
Thus, the CP violating phase $\d^{\rm CKM}$ is expressed by the other matrix elements as 
\begin{align}
\sin \d^{\rm CKM} 
=  { s_{d} c_{d} \over  c_{12} s_{12} } \simeq  {|V_{ts} V_{td}| \over |V_{cb}|^{2} |V_{ud} V_{us}|} 
= 0.953 \, , ~~~ 
 \d^{\rm CKM} \simeq 72.4^{\circ} \, .
\end{align}
Although this CP phase has an error of about 2 \% due to the assumption of $\phi = \pi /2$,  errors in the CKM matrix elements are sufficiently small  because this is normalized by the Jarlskog invariant $J \sim 3 \times 10^{-5}$. 

In this case, the mass matrices $m_{u,d} = U_{u,d} m^{\rm diag}_{u,d} $
reconstructed from $U_{u,d}$ have diagonal reflection symmetries defined as \cite{Yang:2020goc} 
\begin{align}
R \, m_{u}^{*}  = m_{u} \, , ~~~
m_{d}^{*} = m_{d} \, , ~~~ 
R = {\rm diag} (-1 \, , 1 \, , 1) \, . 
\label{DRS}
\end{align}
In this basis, $m_{u,d}$ take the following form,
\begin{align}
m_{u} = 
\begin{pmatrix}
i \, m_{u 11} & i \, m_{u 12} & i \, m_{u 13} \\
m_{u 21} & m_{u 22} & m_{u 23} \\
m_{u 31} & m_{u 32} & m_{u 33} \\
\end{pmatrix} , 
~~~
m_{d} = 
\begin{pmatrix}
m_{d 11} & m_{d 12} & m_{d 13} \\
m_{d 21} & m_{d 22} & m_{d 23} \\
m_{d 31} & m_{d 32} & m_{d 33} \\
\end{pmatrix} , 
\label{12}
\end{align}
with real parameters $m_{u ij}, m_{d ij}$. 

The symmetries defined in this way are clearly distinct from the conventional CP symmetry, because the relative phase cannot be removed by the weak basis transformation (WBT) $q_L' = U_L q_L$  \cite{Branco:1999nb,Branco:2007nn}, which leaves the CKM matrix $V_{\rm CKM} = U_u^\dagger U_d$ invariant. 
If one attempts to remove the phase in Eq.~(\ref{12}), the phase is merely transferred to the down-type quarks as $m_u^* = m_u$ and $R m_d^{*} = m_d$. 
Conversely,  from an arbitrary basis, an appropriate WBT 
yields a basis where $m_{u,d}$ almost satisfy the CP symmetries given in Eq.~(\ref{DRS}). 

Although the original paper defined the vector-like symmetry $R \, m_{u,\n}^{*} \, R = m_{u,\n}$ and $m_{d,e}^{*} = m_{d,e}$, this chiral form seems to be more essential. 
By considering unphysical basis transformations of right-handed fields $u_{R}' = u_{R} V_{u}$ and $d_{R}' = d_{R} V_{R}$, 
GCPs for $\tilde m_{u,d} = U_{u,d} m^{\rm diag}_{u,d} V_{u,d}^{\dg}$ are found to be
\begin{align}
R \, \tilde m_{u}^{*} \, R_{u}  =  \tilde m_{u} \, , ~~~ \tilde m_{d}^{*} = \tilde m_{d} R_{d} \, , ~~~
R_{u,d} \equiv V_{u,d}^{*} V_{u,d}^{\dg} \, .
\end{align}
The chiral GCPs~(\ref{DRS}) are preserved if the field redefinitions $V_{u,d}$ are real $V_{u,d}^{*} = V_{u,d}$.
Since such symmetries are often realized in the context of zero-textures \cite{Shin:1985cg, Shin:1985vi, Lehmann:1995br, Kang:1997uv, Mondragon:1998gy, Fritzsch:1999im, Fritzsch:2002ga, Xing:2003yj, Xing:2009eg, Barranco:2010we, Xing:2015sva, Fritzsch:2021ipb, Awasthi:2022nbi}, 
we do not discuss the realization of the symmetries here. 

\section{Analytic structure of $U_{\rm MNS}$ and physical consequences}

In this section, we analyze the MNS matrix with DRS and its physical consequences. 
Under these symmetries, the mass matrix of charged leptons $m_{e}$ is real and that of Majorana neutrinos $m_{\n}$ has the following form; 
\begin{align}
m_{e}^{*} = m_{e} \, , ~~ 
m_{\n} = 
\begin{pmatrix}
m_{11} & i m_{12} & i m_{13}  \\
i m_{12} & m_{22} & m_{23} \\
i m_{13} & m_{23} & m_{33} \\
\end{pmatrix} \, , 
\end{align}
with real matrix elements $m_{ij}$. 

Since the singular value decomposition of a real matrix is done 
 by a real orthogonal matrix $O_{f}$, the MNS matrix $U$ is 
\begin{align}
U &= 
O_{e}^{T}
 \Diag{i}{1}{1} 
O_{\n} P, \label{UVn0} \\
~~~ 
O_{\n} P &= 
\begin{pmatrix}
 1 & 0 & 0 \\
 0 & c_{\n} & s_{\n} \\
 0 & - s_{\n} & c_{\n} \\
\end{pmatrix}
\begin{pmatrix}
c_{13} & 0 & s_{13} \\
 0 & 1 & 0 \\
- s_{13} & 0 & c_{13} \\
\end{pmatrix}
\begin{pmatrix}
c_{12} & s_{12} & 0 \\
 - s_{12} & c_{12} & 0 \\
 0 & 0 & 1 \\
\end{pmatrix} 
\Diag{e^{i \phi_{1}}}{e^{i \phi_{2}}}{e^{i \phi_{3}}} . 
\label{UVn}
\end{align}
Here, phases $\phi_{i} = 0$ or $\pi /2$ in the phase matrix $P$ originate from the redefinition of  positive or negative mass singular values $\pm m_{\n i}$ after a real diagonalization by $O_{\n}$.

For a hierarchical $m_{e}$ (i.e., $|(m_{e})_{33}| \simeq m_{\t} \gg |(m_{e})_{1i , j1}|$), 
the 1-3 mixing of $m_{e}$ is perturbatively evaluated as $s_{13}^{e} \simeq |(m_{e})_{13} / (m_{e})_{33}|$ and is suppressed by the mass of tau lepton $m_{\t}$. 
This hierarchy is explained by approximate chiral symmetries of the first generation.
The matrix $m_{e}$ has $U(1)_{1L} \times U(1)_{1R}$ symmetry in the limit of the lightest charged lepton is massless, i.e., $m_{e1} \to 0$; 
\begin{align}
m_{e}^{\rm diag} = 
\Diag{0}{m_{e2}}{m_{e3}}  \, ,  ~~~
\Diag{e^{i \a_{L}}}{1}{1}
m_{e}^{\rm diag}
\Diag{e^{i \a_{R}}}{1}{1} = m_{e}^{\rm diag} \, .
\end{align}
If the unitary matrices diagonalizing $m_{e}$ have small mixings, 
approximate chiral symmetries roughly constrain the texture of $m_{e}$ as
\begin{align}
m_{e} \simeq (m_{e})_{33}
\begin{pmatrix}
\e_{L} \e_{R} & \e_{L} & \e_{L} \\
\e_{R} & * & * \\
\e_{R} & * & 1
\end{pmatrix} , 
\end{align}
where $\e_{L,R}$ are breaking parameters of the $U(1)_{1(L,R)}$ symmetry, 
and $*$ represents matrix elements from which $O(1)$ coefficients are omitted.

The situation is divided into three classes depending on the magnitude of $\e_{L,R}$. 
\begin{enumerate}
\item $|\e_{R}| \sim 1 \gg |\e_{L}|$: 
The 1-3 mixing is approximately evaluated as 
\begin{align}
s_{13}^{e} \simeq |\e_{L}| \simeq m_{e1} / m_{e3} = 0.0003 \, , ~~~
\end{align}
and safely ignored. 

\item $|\e_{L}| \sim |\e_{R}|$: 
Similarly, 
\begin{align}
s_{13}^{e} \simeq |\e_{L}| \simeq \sqrt{m_{e1} / m_{e3}} = 0.017 \, .
\end{align}
Therefore, its contribution is at most about a $\pm$10\% change in $s_{13}$. 

\item $|\e_{L}| \gg |\e_{R}|$: 
Although the approximation cannot be used, the unitary matrix for left-handed fields has large mixings and the assumption does not hold.  
With the corresponding CKM matrix element $|V_{ub}| \simeq 0.003$ in mind, 
this situation seems difficult to realize in a unified theory without fine-tunings. 
\end{enumerate}
Thus, this approximation is justified in a wide range of parameters.

Under the approximation that the 1-3 mixing of $m_{e}$ is negligible, 
a combination of the 2-3 mixings of $O_{\n}$ and $O_{e}$ yields
a representation of the MNS matrix as 
\begin{align}
U = 
\begin{pmatrix}
c_{e} & s_{e}& 0 \\
- s_{e} & c_{e} & 0 \\
0 & 0 & 1
\end{pmatrix} 
\begin{pmatrix}
i & 0 & 0 \\
 0 & c_{23} & s_{23} \\
 0 & - s_{23} & c_{23} \\
\end{pmatrix}
\begin{pmatrix}
c_{13} & 0 & s_{13} \\
 0 & 1 & 0 \\
- s_{13} & 0 & c_{13} \\
\end{pmatrix}
\begin{pmatrix}
c_{12} & s_{12} & 0 \\
 - s_{12} & c_{12} & 0 \\
 0 & 0 & 1 \\
\end{pmatrix} \Diag{e^{i \phi_{1}}}{e^{i \phi_{2}}}{e^{i \phi_{3}}}
\, .
\label{MNSmatrix}
\end{align}

We eliminate non-physical signs from several sign degrees of freedom for mixing parameters $c_{ij}$ and $s_{ij}$. 
First, the sign of the phase $i$ can be fixed to positive because it is absorbed to $s_{e}$.
Based on the Kobayashi-Maskawa theory \cite{Kobayashi:1973fv}, 
five signs can be made positive by redefinition of phases of six leptons, 
in a similar way to the standard PDG parameterization. 
For the later convenience, we choose the signs of $c_{12}, c_{13}, c_{23}$ and $s_{13}, s_{23}$ to be positive. 
The sign of $c_{e}$ is also fixed because it is changed by multiplying diag $(-1, -1, 1)$ from the left without changing the sign of $s_{ij}$ and $c_{ij}$. 
In addition, two signs are determined from the signs of $\cos \d$ and $\sin \d$ for the Dirac phase $\d$. 
From the following calculation, we choose those of $s_{12}$ and $s_{e}$. 

The Dirac phase is evaluated as the first observable.
%
The standard PDG parameterization is given by \cite{ParticleDataGroup:2018ovx} 
\begin{align}
U_{}^{\rm PDG} &= 
\begin{pmatrix}
c_{12} c_{13} & s_{12} c_{13} & s_{13} e^{-i\d} \\
-s_{12} c_{23} - c_{12} s_{23} s_{13}  e^{i \d} & c_{12} c_{23} - s_{12} s_{23} s_{13} e^{i \d} & s_{23} c_{13} \\
s_{12} s_{23} - c_{12} c_{23} s_{13} e^{i \d} & -c_{12} s_{23} - s_{12} c_{23} s_{13} e^{ i \d} & c_{23} c_{13}
\end{pmatrix} \nn \\
& \times {\rm diag} (1 ,  e^{ i \a_{2} / 2} , e^{ i \a_{3} / 2}) \, .
\label{PDG}
\end{align}
Note that the phases $\phi_{i}$ in $P$ have a potential to set the Majorana phases $\a_{i}$ to $\a_{i} + \pi$ and will be discussed again with the effective mass of double beta decay. 

By comparing the absolute values of the third column of Eq.~(\ref{MNSmatrix}) 
and (\ref{PDG}), $s_{13}$ and $s_{23}$ should satisfy
\begin{align}
| s_{e} c_{13} s_{23} + i c_{e} s_{13} |^{2} & =  (s_{13}^{\rm PDG})^{2} \, , \\
| c_{e} c_{13} s_{23} - i s_{e} s_{13}|^{2} & =  (s_{23}^{\rm PDG} c_{13}^{\rm PDG})^{2} \, ,
\end{align}
and we obtain the following solutions; 
\begin{align}
s_{23} = \frac{\sqrt{(s_{e} s_{13}^{\text{PDG}})^{2} - (c_{e} c_{13}^{\text{PDG}} s_{23}^{\text{PDG}} )^{2} } }{\sqrt{-c_{e}^4 + (c_{e} s_{13}^{\text{PDG}})^{2} - (s_{e} c_{13}^{\text{PDG}} s_{23}^{\text{PDG}})^{2} + s_{e}^4}} \, ,
~~~ 
s_{13} = \frac{\sqrt{ (c_{e} s_{13}^{\text{PDG}})^{2} - (s_{e} c_{13}^{\text{PDG}} s_{23}^{\text{PDG}})^{2} } }{\sqrt{c_{e}^4-s_{e}^4}}  \, . 
\label{s2313}
\end{align}
Similarly, the condition for $s_{12}$ is determined from $|U_{12}|^{2} = |U_{12}^{\rm PDG}|^{2}$, 
\begin{align}
|s_{e} (c_{12} c_{23} - s_{12} s_{13} s_{23})+i s_{12} c_{e} c_{13}|^{2} = (s_{12}^{\rm PDG} c_{13}^{\rm PDG})^{2} \, . \label{13}
\end{align}
Solving $s_{12}$ from Eqs.~(\ref{s2313}) and (\ref{13}), we obtain four solutions. 
Among them, only two solutions are physically inequivalent, and they differ in the sign of $\cos \d$.
Since the sign of $\cos \d$ and $s_{12}$ are almost the same as a result of drawing the plots (Figure 1), 
experiments favor the solution of $s_{12} < 0$. 
Also, 
since the condition (\ref{13}) is a function of $s_{e}^{2}$ and $c_{e}^{2}$, 
the solutions of $s_{12}$ is independent of signs of $s_{e}$ and $c_{e}$.

As input values, we use the latest global fit 
without Super-Kamiokande (SK) in the normal hierarchy (NH) \cite{Gonzalez-Garcia:2021dve}, 
\begin{align}
\sin^{2} \th_{12}^{\rm PDG} = 0.304 \, , ~~~ \sin^{2} \th_{23}^{\rm PDG} = 0.573 \, , ~~~ \sin^{2} \th_{13}^{\rm PDG} = 0.0222 \, ,
\end{align}
because the values of inverted hierarchy (IH) with or without SK are close to these values.  
Although the inclusion of the SK data makes $s_{23}^{\rm PDG}$ about $0.1$ smaller for NH, 
the qualitative behavior in the following discussion remains the same.

From these conditions, $\sin \d$ and $\cos \d$ are expressed as functions of $s_{e}$ and some sign degrees of freedom.
The cosine of $\d$ is given by
\begin{align}
\cos \d & =  { |U_{22}^{\rm PDG}|^{2} - (s_{12}^{\rm PDG} s_{13}^{\rm PDG} s_{23}^{\rm PDG})^{2} - (c_{12}^{\rm PDG} c_{23}^{\rm PDG})^{2} \over - 2 s_{12}^{\rm PDG} s_{13}^{\rm PDG} s_{23}^{\rm PDG} c_{12}^{\rm PDG} c_{23}^{\rm PDG} }    \\
&= {|U_{22}|^{2} (1 - |U_{13}|^{2})^{2} - |U_{13}|^{2} |U_{12}|^{2} |U_{23}|^{2} - |U_{11}|^{2} |U_{33}|^{2} \over - 2 |U_{13}| |U_{12}| |U_{23}| |U_{11}| |U_{33}|} \, . 
\end{align}
Also, $\sin \d$ is evaluated from the Jarlskog invariant \cite{Jarlskog:1985ht}; 
\begin{align}
J = - {\rm Im} \,  [U_{\m 3} U_{\t 2} U_{\m 2}^{*} U_{\t 3}^{*}] 
&= \sin \d \, s_{12}^{\rm PDG} c_{12}^{\rm PDG} s_{13}^{\rm PDG} (c_{13}^{\rm PDG})^{2} s_{23}^{\rm PDG}  c_{23}^{\rm PDG} \\
& =  c_{13} c_{23} c_{e} s_{e} (c_{12} s_{23} + c_{23} s_{12} s_{13}) (s_{12} s_{23} - c_{12} c_{23} s_{13})  \, . 
 \label{exp}
\end{align}
It predicts a proportional relationship between $\sin \d$ and $s_{e}$. When $s_{13}$ is small, the invariant roughly leads to
\begin{align}
J \simeq c_{12} c_{23} c_{e} s_{12} s_{23}^2 s_{e} \, , ~~
\sin \d \simeq - { c_{e}  s_{e}  s_{23} \over  s_{13}^{\rm PDG} (c_{13}^{\rm PDG})^{2}  } 
\simeq \pm \, 5 \, s_{e} \, , ~~
\To ~~ \cos \d \simeq \pm \sqrt{1 - 25 s_{e}^{2}} \, .
\end{align}
Then the sign of $s_{e} c_{e}$ and $\sin \d$ are opposite. 
This minus sign comes from the choice $s_{12} \simeq - s_{12}^{\rm PDG}$. 

Figure 1 shows plots of $\cos \d$ and $\sin \d$ expressed as functions of $s_{e}$. 
In the plot of $\cos \d$, the red and green lines correspond to $s_{12} > 0$ and $s_{12} < 0$. 
Since the signs are approximately equal (${\rm sign} (\cos \d) \simeq {\rm sign} (s_{12})$) in the parameter regions, experiments favor $s_{12} < 0$.
For $\sin \d$, the color of the lines depends on the sign of $c_{e}$. 
\begin{figure}[t]
\begin{center}
\begin{tabular}{cc}
 \includegraphics[width=7cm]{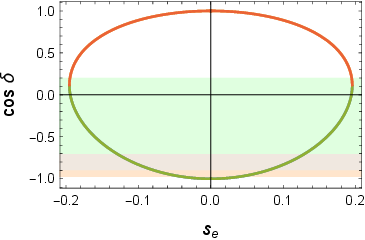} ~&~ 
 \includegraphics[width=7cm]{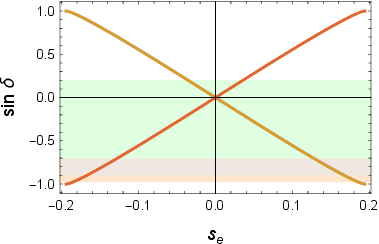} 
\end{tabular}
\caption{ Plots of $\cos \d$ and $\sin \d$ expressed as functions of $s_{e}$. 
The bright green and orange regions represent the $1 \s$ regions of NH and IH (without SK). 
In the plot of $\cos \d$, the red and green lines correspond to $s_{12} > 0$ and $s_{12} < 0$. 
For $\sin \d$, the color of the lines depends on the sign of $c_{e}$. 
}
\end{center}
\end{figure}
The parameter $|\sin \d|$ becomes maximal around $|s_{e}| \simeq 0.2$.
In regions where $s_{e}$ is larger than this value,  $\sin \d$ and $s_{13}$ become complex numbers and have no real solution.  This is due to the following reasons.
In the range of $|s_{e}| \lesssim 0.2$, $s_{e}$ can be regarded as a perturbation. 
From Eq.~(\ref{s2313}), 
$s_{12}$ and $s_{23}$ are only affected by the first and second-order perturbation of $s_{e}$ and $s_{13}$. 
Thus the parameters $s_{12}$ and $s_{23}$ are approximately equal to those of PDG and 
$s_{13}$ is constrained as 
\begin{align}
s_{23} \simeq s_{23}^{\rm PDG} \, ,  ~~~ 
s_{12} \simeq - s_{12}^{\rm PDG} \, , ~~~
s_{13} \simeq  \sqrt{ (s_{13}^{\text{PDG}})^{2} - (s_{e} s_{23}^{\text{PDG}})^{2} } \, . 
\end{align}
Since the maximum value of $s_{e}$ in this range is realized by $s_{13} = 0$, 
\begin{align}
s_{e}^{\rm max} \simeq {s_{13}^{\rm PDG} \over s_{23}^{\rm PDG}} \simeq 0.196 \, . 
\end{align}
In this limit $s_{13} \to 0$, the MNS matrix also results in the representation of Eq.~(\ref{VCKM1}). 

There exist other solutions with $s_{e} \simeq \pm 1$. 
However, since these solutions imply that the eigenstates of the charged leptons $e$ and $\m$ are interchanged by diagonalization, it is excluded from a point of view of the natural mass matrix \cite{Peccei:1995fg}.

\subsection{Majorana phases and effective mass of double beta decay}

A similar analysis is performed for the Majorana phases. 
These phases are evaluated from the following quantities \cite{Branco:1986gr, Jenkins:2007ip, Branco:2011zb}; 
\begin{align}
I_{1} & = {\rm Im} \, [U_{e2}^{2} U_{e1}^{*2} / |U_{e2} U_{e1}|^{2}] =
 \sin \a_{2},  
\label{I1} \\
I_{2} & = {\rm Im} \, [ U_{e3}^{2} U_{e1}^{*2} / |U_{e3} U_{e1}|^{2}] =
\sin (\a_{3} - 2 \d) . 
\label{I2}
\end{align}
Expansions of $\sin \a_{2,3}$ for small $s_{e}$ are respectively
\begin{align}
\sin \a_{2}^{0} \simeq -  \frac{2  s_{e} c_{23} }{ c_{e} c_{12} s_{12} } \simeq + 3 {s_{e} \over c_{e}} \, , ~~~ 
\sin \a_{3}^{0} \simeq -  \frac{2  s_{e} c_{e} c_{12} c_{23} }{ c_{13} s_{12} } \simeq + 2  {s_{e}  c_{e}} \, . 
 \end{align}
Furthermore, $\sin \a_{i}$ has degrees of freedom $\phi_{i} = 0 $ or $\pi / 2$ in Eq.~(\ref{UVn}) due to the signs of singular values;
\begin{align}
\a_{2} = \a_{2}^{0} + 2(\phi_{2} - \phi_{1}) \, , ~~~
\a_{3} = \a_{3}^{0} + 2(\phi_{3} - \phi_{1}) \, .
\end{align}
Ratios of the CP-violating phases are evaluated as 
\begin{align}
\abs {\sin \a_{2} \over \sin \d} & \simeq  
\abs{2 s_{13} c_{13}^{2} c_{23} \over c_{e}^{2} c_{12} s_{12} s_{23}} \simeq 
 {3 \sqrt 2 s_{13} } \simeq  0.6 \, , \\ 
\abs {\sin \a_{3} \over \sin \d} & \simeq 
\abs{2   c_{12} c_{23}  c_{13} s_{13} \over s_{12}  s_{23} } \simeq  2 \sqrt{2} s_{13} \simeq 0.4 \, .
\end{align}
As a result, we obtain correlations between the mixing angle $s_{e}$ and CP-violating phases 
\begin{align}
|\sin \delta|  \simeq 1.6 |\sin \alpha_{2}| \simeq 2.5 |\sin \alpha_{3}| \simeq 5 |s_{e}| \, . 
\end{align}
Since the MNS matrix~(\ref{MNSmatrix}) is CP-symmetric in the limit of $s_{e} \to 0$, 
it is a natural consequence that these CP phases have such correlations.

Since the three CP phases are written by functions of $s_{e}$, 
we can represent the effective mass of the double beta decay  $m_{ee}$ by $s_{e} \simeq \pm \sin \delta / 5$ and the lightest neutrino mass $m_{1 \, \rm or \rm \, 3}$; 
\begin{align}
m_{ee} = U_{e1}^{2} m_{1} + U_{e2}^{2} m_{2} + U_{e3} m_{3}^{2} \, . 
\end{align}
The phases $e^{2 i \phi_{i}}$ in Eq.~(\ref{UVn}) are fixed to $\pm 1$ by the GCPs. 
Due to the overall phase transformation, we do not lose generality by setting $\phi_{3} = 0$. 
Therefore, there exist four types of solutions ($e^{2 i \phi_{1}} = \pm 1, e^{2 i \phi_{2}} = \pm 1$) for the case of NH and IH, respectively.
The values of the mass differences are also taken from the latest global fit (without SK)  \cite{Gonzalez-Garcia:2021dve}, 
\begin{align}
\D m^{2}_{21} = 74.2 \meV^{2} \, , ~~ \D m^{2}_{31} = 2515 \meV^{2} \, , ~~ 
\D m^{2}_{32} = - 2498 \meV^{2} \, . 
\end{align}
\begin{figure}[t]
\begin{center}
 \includegraphics[width=16cm]{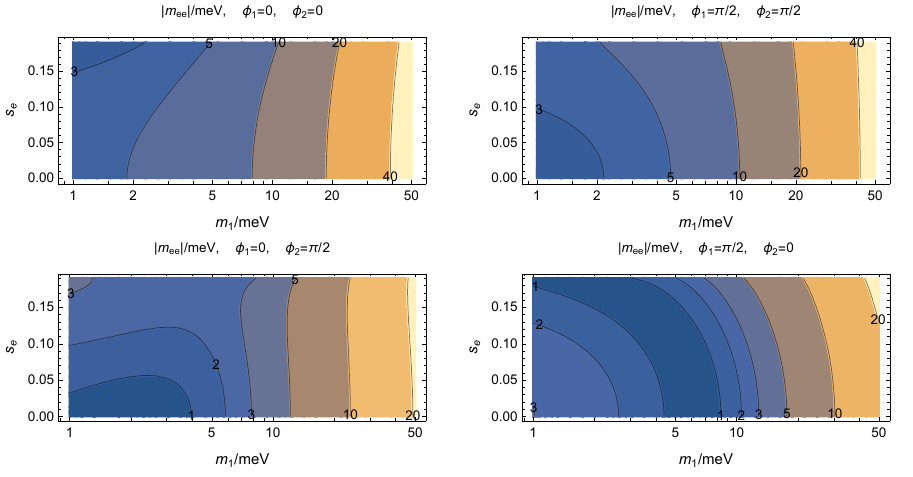} \\
 \includegraphics[width=16cm]{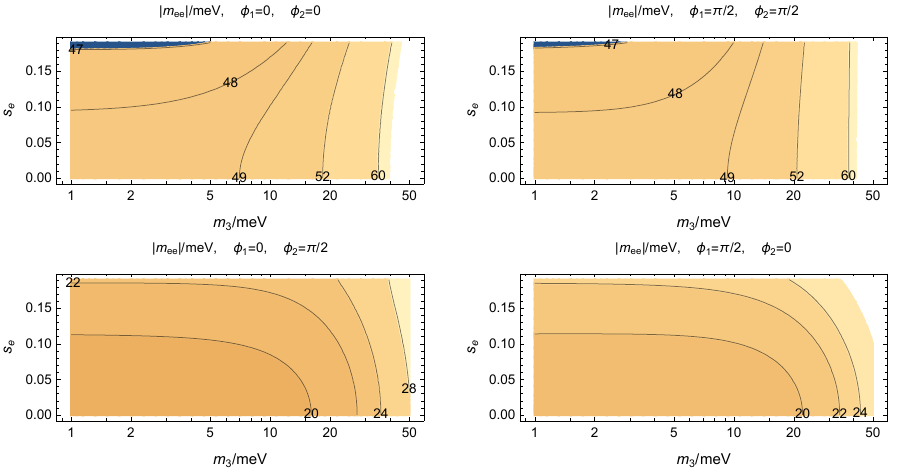}
\caption{Plots of the absolute value  of the effective mass of double beta decay $|m_{ee}|$ for NH and IH as functions of $s_{e}$, $m_{1}$, and $\phi_{1,2}$.}
\end{center}
\label{}
\end{figure}
Figs.~2 shows the absolute value of the effective mass $|m_{ee}|$ as a function of $s_{e}$ and $m_{1}$ in the NH and IH cases. The range of $s_{e}$ is set to $0 < s_{e} < 0.2$ because the results do not depend on the sign of $s_{e}$.
These behaviors can be understood as follows. 
Since the three CP phases are small for sufficiently small $s_{e}$, the mass $m_{ee}$ is approximately 
\begin{align}
m_{ee} \simeq m_{1} c_{13}^{2} c_{12}^{2} e^{ 2 i \phi_{1}} + m_{2} c_{13}^{2} s_{12}^{2} e^{2 i \phi_{2}}
+ m_{3} s_{13}^{2} \, . 
\end{align}
Thus, the sign between $e^{ i \phi_{1}}$ and $e^{ i \phi_{2}}$ causes constructive and destructive interferences in $m_{ee}$.
This result corresponds to near the maximum or minimum value in the whole parameter region, 
because the CP phases are all small and an extremum is chosen by the sign $e^{2 i \phi_{i}}$.

In particular, in the case of IH with $m_{1} \simeq m_{2} \gg m_{3}$, the same sign leads to $|m_{ee}| \simeq m_{2}$, and the different sign leads to $|m_{ee}| \simeq m_{2} / 3$.
The same sign scenarios are suggested to be excluded by the latest limit of the KAMLAND-Zen collaboration $|m_{ee}| \lesssim 36 - 156 \meV$ \cite{KamLAND-Zen:2022tow}.

\section{Summary}

In this paper, we surveyed the analytic structure of the MNS matrix with diagonal reflection symmetries. 
If the mass matrix of charged leptons $m_{e}$ is hierarchical (i.e., $|(m_{e})_{33}| \simeq m_{\t} \gg |(m_{e})_{1i , j1}|$), by neglecting the 1-3 mixing of $m_{e}$, the MNS matrix is represented by four parameters and several sign degrees of freedom.
By substituting the three observed mixing angles $\th_{ij}$ as input parameters, 
the Dirac phase $\d$ and the Majorana phases $\a_{2,3}$ are represented functions of the 1-2 mixing of charged leptons $s_{e}$. 
As a result, we obtain clear correlations between CP-violating phases 
$|\sin \d| \simeq 1.6 |\sin \a_{2}| \simeq 2.5 |\sin \a_{3}|$. 

The effective mass of double beta decay $m_{ee}$ is also displayed as a function of $s_{e}$ and 
the lightest neutrino mass $m_{1 \, \rm or \, 3}$.
Because the generalized CP symmetries restrict the effective mass to near the maximum or minimum value in the whole parameter region, several scenarios are suggested to be excluded by the latest limit of the KAMLAND-Zen collaboration. 

\section*{Acknowledgment}

This study is financially supported 
by JSPS Grants-in-Aid for Scientific Research
No.~JP18H01210 
and MEXT KAKENHI Grant No.~JP18H05543.


\end{document}